\documentclass{aa}  
\usepackage{amsmath,amssymb} 
\usepackage{txfonts}
\usepackage[T1]{fontenc}
\usepackage{afterpage}
\usepackage{amssymb}
\usepackage{natbib} 
\usepackage{wrapfig}
\usepackage{graphicx}
\usepackage{natbib}
\usepackage{amsmath}
\usepackage{color}
\usepackage[normalem]{ ulem }
\usepackage{soul}
\usepackage{txfonts}
\bibpunct{(}{)}{;}{a}{}{,} 
\def\mgb{$^{25}$Mg}

\def\DY{$\Delta$Y}

\newcommand{\ms}{M$_{\odot}$}

\newcommand{\gcc}{$\rm {g \ cm^3}$}

\begin{document}

   \title{Revisiting nucleosynthesis in globular clusters: the case of NGC~2808 and the role of He and K}
\titlerunning{Nucleosynthesis in globular clusters}

   \author{        
          N. Prantzos 
          \inst{1},
          C. Charbonnel
          \inst{2,3},
          C. Iliadis
          \inst{4}
          }

   \institute{$^1$Institut d'Astrophysique de Paris, UMR7095 CNRS, Univ. P. \& M. Curie, 98bis Bd. Arago, 75014 Paris, France \\
     \email{prantzos@iap.fr} \\
   $^2$ Department of Astronomy, University of Geneva, Chemin des Maillettes 51, 1290 Versoix, Switzerland \\
   $^3$ IRAP, UMR 5277, CNRS and Universit\'e de Toulouse, 14, av. E. Belin, F-31400 Toulouse, France \\
   $^4$ Department of Physics \& Astronomy, University of North Carolina at Chapel Hill, Chapel Hill, NC 27599-3255, USA \\
                          }
\date{}


 
  \abstract
   {Motivated by recent reports concerning the observation of limited enrichment in He but excess K in stars of globular clusters, we revisit the H-burning conditions that lead to the chemical properties of multiple stellar populations in these systems.}
   {In particular, we are interested in correlations of He and K with other elements, like O, Na, Al, Mg and Si, reported in stars of NGC~2808 
   }
   {We perform calculations of nucleosynthesis at constant temperature and density, exploring the temperature range of 25 to 200 10$^6$ K (25 to 200 MK), using a detailed nuclear reaction network and the most up-to-date nuclear reaction rates.}
   {We find that Mg is the most sensitive "thermometer" of hydrostatic H-burning conditions, pointing to a temperature range of 70-80 MK for NGC~2808, while He is a lesser - but not negligible - constraint. K can be produced at the levels reported for NGC~2808 at temperatures $>$180 MK and Si at T$>$80 MK. However, in the former temperature range Al and Na are totally destroyed and no correlation can be obtained, in contrast to the reported observations. None of the putative polluter sources proposed so far seem to satisfy the ensemble of nucleosynthesis constraints.}
   {}


   \maketitle
%


\section{Introduction}
\label{sec:Intro}


The discovery of the ubiquitous presence of multiple stellar populations (MSP) in Galactic globular clusters (GC) is considered as one of the major breakthroughs of the past two decades in stellar population studies. It has motivated the massive acquisition of exquisite photometric and spectroscopic data with large ground-based and space surveys to probe the peculiarities of numerous stars in a significant number of GCs. 
On the photometric side, the analysis of GCs color-magnitude diagrams (CMD) 
combining optical, ultraviolet, and infrared HST filters has definitively challenged the classical view of GCs being single isochrones. This method has revealed in different areas of  GCs CMDs the presence of diverse and complex 
sequences\footnote{Some clusters show e.g. two well separated group of stars on the subgiant branch and the red giant branch, while in others these two sequences appear to be merged into a single broad sequence \citep[e.g.][]{Miloneetal17}. that are presumably occupied by MSPs (e.g. \citealt{Miloneetal12a,Miloneetal17,Piotto15,Milone16,Sotoetal17}). This is supported by the fact that stars on the different sequences also exhibit different chemical compositions \citep[e.g.][]{Marinoetal08,Marinoetal16,Yongetal08,Miloneetal12a}. }.
Indeed, multi-objects spectroscopy with VLT has extensively documented the chemical properties of MSPs, revealing that GC stars located in all the regions of the CMD (main-sequence, turnoff, red giant, horizontal and asymptotic giant branches) exhibit large abundance variations of C, N, O, Na, Al, and sometimes Mg that are coupled with anti-correlations (C-N, Na-O, Mg-Al, Li-Na; e.g.  \citealt{Carrettaetal04,Carrettaetal09b,Carretta06,Lindetal09,D'Orazietal10,Gratton12,Wangetal16}), the exact shape and extension of which vary from one cluster to another 
(in particular, Mg depletion - thus the Mg-Al anticorrelation - is seen only in $\sim 40 \%$ of Galactic GCs, and is more extended in the most massive or most metal-poor GCs; \citealt[][and references therein]{Pancinoetal17}).

These abundance patterns have long been recognized to result from hydrogen-burning at high temperature through the CNO-cycle and the NeNa- MgAl-chains \citep{KudryashovTutukov98,DenisenkovDenisenkova90,Langeretal93,Denissenkovetal98,Prantzos06,PCI07}. The fact that the sum C+N+O is observed to be constant from star to star (\citealt{Dickens91,Ivansetal99,Carrettaetal05,CohenMelendez05,Smithetal05,Yongetal08}; note however the cases of NGC~1851, \citealt{Yongetal09}, and M~22, 
\citealt{Marinoetal11M22}) is a clear signature of H-burning. Elements heavier than those affected  by the MgAl chain or by more advanced phases of stellar nucleosynthesis were thought  to be essentially mono-abundant in GCs. However, that idea is challenged by recent observations (see below). 

The spectroscopic identification of MSPs indicates that $\sim 30 \%$ of the GC stars were born with similar chemical properties as their contemporary halo field counterparts (i.e., with typical SNeII patterns); these chemically normal stars are often referred to as first population, or first generation stars (1P or 1G).  The rest $\sim 70 \%$ are considered to be born from a mixture of the pristine proto-cluster gas with hydrogen-burning products ejected by short-lived GC stars that have polluted the GCs in the very early phases of their evolution (\citealt{Ventura01,Prantzos06,Carretta13}; see also \citealt{Charbonnel14} who propose a test on the actual percentage of GC stars born out of polluted material); these are the so-called second population or second generation stars (2P or 2G). 

When using photometric indices to separate the different populations on the RGB, the fraction of 1P stars varies between $\sim 8$ and $\sim 67 \%$ from one GC to another, and it decreases with the GC mass \citep{Miloneetal17}.
So far, the MSP phenomenon has been detected 
in all the Galactic GCs where it has been looked for, 
as well as in some extragalactic GCs close enough for appropriate observations to be carried out\footnote{The only known exception in the Milky Way is Rup~106 \citep{Villanovaetal13}, an old ($\sim$ 12~Gyr) and relatively metal-poor ([Fe/H]=-1.5) low-mass GC that appears to be much less compact than MSP GCs \citep{Krauseetal16}. Other chemical peculiarities of this cluster (i.e., lack of $\alpha$-enhancement) point towards an extragalactic origin. We note that another GC from the Sagittarius elliptical dwarf galaxy, namely Pal 12 does not seem to show the O-Na anticorrelation \citep{Cohen04}.}. It has never been found however in any other stellar populations like open clusters (\citealt{deSilvaetal09,Pancinoetal10,Bragagliaetal12,Bragagliaetal14}), with the possible exception of the old, very massive, and very-metal rich galactic open cluster NGC~6791  \citep{Geisleretal12}. It is intensively searched for in young massive star clusters (YMSC)
and evidence for N-spread from photometry and spectroscopy points for their presence in clusters as young as 2~Gyr (but not in younger clusters, at least not for stars evolved past the main sequence turnoff; \citealt{Bastian2016EAS,Krauseetal16,Martocchiaetal17,Niederhoferetal17I,Niederhoferetal17II,Hollyheadetal17}; Martocchia et al. submitted).

Several enrichment scenarios have been developed to decipher the photometric and 
chemical specificities of  GC MSPs, and in particular to reproduce the well-documented O-Na and Mg-Al anticorrelations. Each of them calls for different stellar sources of hydrostatic hydrogen-burning ashes, implying different pollution modes of the intra-cluster gas and consequently different secondary star formation  routes for the MSPs \citep[for a review see][]{Charbonnel16_ees}. The most commonly invoked polluters are fast rotating massive stars (FRMS; $\geq$ 25~M$_{\odot}$; \citealt{Maeder06,Prantzos06,Decressin07a,Decressin07b,Krauseetal13,Krauseetal16}), massive AGB stars ($\sim$6.5~M$_{\odot}$; e.g. \citealt{Ventura01,Ventura13,Decressin09a,D'ercole12,D'Antonaetal16}),  and intermediate-mass binaries (10 - 20 ~M$_{\odot}$; \citealt{DeMink09}).
Recently, supermassive stars (SMS, $\geq 10^4$~M$_{\odot}$; \citealt{DenissenkovHartwick14,Denissenkovetal15}) appeared as potential important players. Since the invoked H-burning operates in different conditions and at different evolution phases within these possible polluters (on the main sequence for the FRMS and the SMS, during the thermal pulse phase for the AGB), model predictions differ in many ways. Yet, none of the current scenarios is able to reproduce the extreme variety of spectroscopic and photometric signatures of GC self-enrichment (e.g. \citealt{Bastianetal15,Renzinietal15,Charbonnel_iaus316,Krauseetal16}). 

In addition to the well-documented 
O-Na and Al-Mg anticorrelations, 
recent determinations of the abundance of other key chemicals like He, K, Si, and Ar isotopes have opened new theoretical challenges. 
In particular, isochrone fitting of high precision photometric data for a handful of GCs calls for relatively low helium enrichment between $+$0.01 and $+$0.2 in mass fraction ($\Delta$Y)\footnote{Assuming a typical He mass fraction Y of 0.25 for the proto-cluster gas, this means that the some GC stars were born with Y of maximum 0.45.} among MSPs, the highest values being derived for those GCs that exhibit the most extended O-Na anticorrelations and are among the most massive within the Milky Way \citep[e.g.][]{Piotto07,Bussoetal07,Caloi07,King12,Miloneetal12,Milone13,Milone15,Leeetal13NGC2419,Bellinietal17}.
The lowest values for He enrichment are hard to reconcile with the  FRMS and the AGB scenarios which both require a minimum $\Delta$Y of $\sim$0.13 - 0.15 to reproduce the bulk abundance variations, while higher values are needed to account for the most extreme part of the O-Na anticorrelation (see \citealt{Bastianetal15}, \citealt{Chantereau16}, and Chantereau et al. 2017 for extended discussions of the various issues raised by the He constraints).
On the other hand, the surprising star to star abundance variations of K reported for a couple of GCs \citep{Cohenetal11,CohenKirby12,Mucciarellietal12} might still be interpreted in the framework of the H-burning scenario, but at much higher temperature than previously thought \citep{Venturaetal16_K,Iliadisetal16}. 

In this context, it is timely to revisit the physical nucleosynthesis conditions that shaped the chemical properties of GC MSPs, 
taking into account the new constraints from He, Si, Ar, and K abundance determination in GCs. This is the aim of the present paper, where we focus on the case of NGC~2808 for which data exist for all these elements. 
We describe the corresponding large body of observational data in \S~\ref{sec:Observations_NGC2808}. We then follow the  procedure developed by \citet[][hereafter PCI07]{PCI07} and extend it to nucleosynthesis calculations over a much broader range of temperature and a larger nuclear network with updated nuclear reactions. All the details on the method, assumptions, and nuclear input physics are given in \S~\ref{sec:Nucleo} where we present the results of our calculations for H-burning at constant temperature. In \S~\ref{sec:CompNGC2808}
we use the nucleosynthesis calculations to explore the dilution factors between H-processed and unprocessed material that are required to reproduce the C-N, O-Na, Mg-Al, anticorrelations as well as the behaviour of the Mg and K isotopes observed in NGC~2808. 
In \S~\ref{sec:Polluters} we discuss the astrophysical sites where the required nucleosynthesis conditions are expected to be fulfilled and derive constraints on the corresponding stellar polluters before concluding in \S~\ref{section:summary}.


\section{Abundance patterns in NGC~2808}    
\label{sec:Observations_NGC2808}

Here we summarize the observational data from the literature regarding the chemical composition of the stellar populations in NGC~2808 which we will use in \S~\ref{sec:Nucleo} and \ref{sec:CompNGC2808} to constrain the H-burning temperatures and the nature of the GC polluters. The numbers are given in  Table~\ref{tab:variationsabundancesNGC2808} 
and the observational data are shown in Fig.~\ref{Fig:CompDataAllElements}. We also recall what are the peculiarities of NGC~2808 for certain elements with respect to other Galactic GCs. 

\subsection{Fe, O, Na, Mg, Al, Si}    
\label{subsec:FeONaMgAl}

NGC~2808 is a moderately metal-poor GC, with a mean [Fe/H] of -1.1 and no significant intrinsic scatter in Fe, Fe-group and $\alpha$-elements \citep[][and references therein]{Carretta15,Wangetal16}.
It is one of the brightest \citep[M$_V$=-9.39;][]{Harris10}, most massive \citep{Kimmig_etal15}, and most compact \citep{Krauseetal16} Galactic GCs. This is an important point because the present-day GC mass and compactness\footnote{The compactness index is usually defined as (M$_*$/$10^5$)/(r$_h$/pc), with M$_*$ the stellar mass and r$_h$ the half-mass radius of a GC. Additionally, \citet{Pancinoetal17} find that the extent of the Mg-Al anticorrelation also depends on GC [Fe/H] content.} appear to be the main parameters driving the extent of the abundance variations among GC MSPs \citep[e.g.][]{Carrettaetal10,Krauseetal16,Miloneetal17}.  The Na-O and Mg-Al anticorrelations in this cluster are indeed very extended, with variations in [Na/Fe], [O/Fe], [Mg/Fe], and [Al/Fe] reaching $+0.7$, $-1.3$, $-0.5$, and $+1.1$~dex respectively \citep{Carretta15}. A Mg-Si anticorrelation could also be detected among RGB stars in this GC, with [Si/Fe] variations of $+0.15$~dex among the most Mg-depleted stars \citep{Carretta15}.

\subsection{K}    
\label{subsec:K}
\citet{Mucciarellietal15} derived [K/Fe] abundance ratios in NGC~2808 for a large sample (119) of RGB stars with large O, Na, Mg, and Al variations. They found that the small subsample (four stars) of Mg-deficient stars show K enrichment by $+0.3$~dex compared to the Mg-normal stars, indicating the existence of a Mg-K anticorrelation in this cluster. 

A similar pattern, although of larger amplitude, was found in another massive GC, namely NGC~2419, with [K/Fe] enhancement derivations that very between 0.78~dex \citep{CohenKirby12} and 2~dex \citep{Mucciarellietal12} in the Mg-depleted subpopulation. So far however, no significant intrinsic spread in [K/Fe] has been found in any other GC \citep{Carrettaetal13Potassium}, although marginally significant K-O anticorrelation in NGC~104 (47~Tuc) and NGC~6809 and K-Na correlation in NGC~104 and NGC~6809 can not be excluded \citep[][but see \citealt{Cerniauskasetal17} who finds no star-to-star abundance variation of K in NGC~104]{Mucciarellietal17}.  

\subsection{He}    
\label{subsec:He}
Relatively large internal helium abundance variations have been derived among NGC~2808 MSPs with indirect methods based on the interpretation of color distributions in different parts of the CMD. $\Delta$Y values between 
0.13 and 0.2 were inferred to explain the horizontal branch (HB) morphology and the width of the main sequence \citep{D'AntonaCaloi04,D'Antona05,Leeetal05He,Piotto07,Dalessandro11,Miloneetal12}. Values between 0.13 and 
0.19 were derived from the analysis of the RGB bump luminosity function \citep{Bragagliaetal10a,Miloneetal15a}. 

Challenging spectroscopic He determinations have also been attempted. A value of $\Delta$Y no less than 0.17 was obtained through the comparison of the He 10830 {\AA} chromospheric line strengths in two red giant stars exhibiting a difference in [Na/Fe] of 0.52~dex \citep{Pasquinietal11}.  
Direct spectroscopic measurements of non-local thermodynamic equilibrium He abundances indicate $\Delta$Y values of 
$0.09 \pm 0.06$ for a subset of blue horizontal branch (HB) stars in NGC~2808 \citep{Marino14}; for these objects however the strong effects of atomic diffusion may blur the interpretation of the data.

We note that the other GCs for which similarly high He variations have been derived are also among the most massive Galactic objects, namely $\Omega$~Cen \citep[NGC~5139, $\Delta$Y $\geq$ 0.14 -- 0.17 needed to explain the blue main sequence; e.g.,][]{Norris04,Piotto05OmCen,Moehleretal07,DupreeAvrett13,Bellinietal17}, NGC~6441 \citep[$\Delta$Y $\sim$ 0.15 needed to reproduce the blue HB;][]{Caloi07}, and NGC~2419 ($\Delta$Y $\sim$ 0.11 - 0.18 
needed to reproduce the blue HB, \citealt{DiCriscienzoetal11,DiCriscienzoetal15}; $\Delta$Y=0.19 from Subaru narrowband photometry of the RGB, \citealt{Leeetal13NGC2419}) which also shows K abundance spread (\S~\ref{subsec:K}).

For other GCs analysed with multiwalength HST photometry, much lower He abundance spread ($\Delta$Y below 0.03) have been inferred to fit the various patterns in the CMD ($\Delta$Y = 0.013 $\pm$ 0.001 for NGC~288, \citealt{Piotto13}; 0.01 for NGC~6397, \citealt{Miloneetal12ngc6397};
0.029 $\pm$ 0.006 for NGC~6352, \citealt{Nardielloetal15}; 0.08 $\pm$ 0.01 for NGC~6266, \citealt{Milone15}). Similar results were obtained through multivariate analysis of the HB morphology for a large sample of GCs observed with HST \citep{Miloneetal14}, for which He was derived with the R-parameter method \citep{Iben6868Nature,Salaris04,Gratton10}.

\begin{table}[h]
\begin{center}
\caption{Initial and extreme abundance values derived for various elements in NGC~2808 (\citealt{Carretta14} for Al; \citealt{Carretta15} for O, Na, Mg, Si; \citealt{Mucciarellietal15} for K) and the corresponding spreads 
}
\begin{tabular}{c c c c}
\hline
 & Init.[X/Fe] & Extr. [X/Fe] & $\Delta$ [X/Fe]\\
O & 0.4 & -0.9 & -1.3 \\ 
Na & -0.1  & 0.6 & + 0.7 \\ 
Mg & 0.4 & -0.1 & -0.5 \\ 
Al & 0 & 1.1 & +1.1\\ 
Si & 0.25 & 0.4 & +0.15 \\ 
K & -0.05 & 0.25 & + 0.3 \\ 
\hline
 &  &  & $\Delta$ Y \\
He & & & 0.13 - 0.2 \\
\hline
\end{tabular}
\label{tab:variationsabundancesNGC2808}
\end{center}
\end{table}


\section{Hydrogen-burning nucleosynthesis in hydrostatic conditions}    
\label{sec:Nucleo}

\subsection{Nuclear reaction network and initial composition}
\label{sub:Network}

We perform H-burning nucleosynthesis calculations at constant temperature T and density $\rho$ within a range that corresponds to hydrostatic burning in the cores of massive and super-massive stars or the H-shells of AGBs. 
The adopted reaction network consists of 213 nuclear species, from protons and neutrons up  to $^{55}$Cr. Thermonuclear reaction rates are adopted from STARLIB 3 \citep{Sallaskaetal13} in tabular form, in the temperature range 10 MK to 10 GK. 
It also includes stellar weak interaction rates depending on both temperature and density.  The network and the reaction rates are presented in  detail in \citet{Iliadisetal16}.

Potassium ($^{39}$K) is made through the reaction chain:
$^{36}$Ar(p,$\gamma$)$^{37}$K($\beta^+$)$^{37}$Ar(p,$\gamma$)$^{38}$K($\beta^+$)$^{38}$Ar(p,$\gamma$)$^{39}$K and it is destroyed  mainly through
$^{39}$K(p,$\gamma$)$^{40}$Ca.
As emphasized in \citet{Iliadisetal16},  
"most of the important reaction rates
for studying hydrogen burning in globular clusters are based on experimental nuclear physics
information and provide a reliable foundation for robust predictions". This concerns, in particular,
some of the important reactions involved in the production of potassium, namely $^{36}$Ar(p,$\gamma$)$^{37}$K,  $^{38}$Ar(p,$\gamma$)$^{39}$K, $^{39}$K(p,$\gamma$)$^{40}$Ca.
In contrast, a few  other reactions in that region, like 
$^{37}$Cl(p,$\gamma$)$^{38}$Ar, $^{37}$ Ar(p,$\gamma$)$^{38}$K, and 
$^{39}$K(p,$\alpha$)$^{36}$Ar
have no experimentally measured reaction rates and statistical calculations with the TALYS code have been adopted.

For the initial composition of the mixture we adopt the one of field halo stars of the same metallicity, [Fe/H]=--1.2, as NGC~2808.
The abundances of the light isotopes are those resulting from Big Bang nucleosynthesis, with a $^7$Li abundance of log(Li/H)+12=2.65 (after BBN+WMAP).

The abundances of ONaMgAl are those suggested by the observations of the most O-rich stars in NGC~2808, 
namely those having [O/Fe]$\sim$0.4, while for C and N we adopt [X/Fe]=0. We adopt here the values of \cite{Lodders2010} as our reference solar system abundances. For the initial isotopic abundances, we rely on galactic chemical evolution 
calculations. We update the calculations of 
\citet{GoswamiPrantzos00} for the chemical evolution of the Galactic halo,
by using more recent ingredients:
a set of metallicity dependent stellar yields  from \citet{Nomotoetal13}  from massive stars and \citet{Karakas10} for intermediate-mass stars, as well as empirical prescription for the rate of SNIa (see Appendix C in \citet{Kubryk2015a}  for a detailed description of the updated chemical evolution ingredients). The chemical evolution model allows us to predict the abundances of all isotopes at a metallicity of  [Fe/H]$\sim$--1 and we adopt those values here as initial ones (i.e. corresponding to the  composition of the gas of the globular cluster and of its 1st generation stars) for our nucleosynthesis and mixing calculations.

\subsection{Evolution of the composition during H-burning at constant temperature}
\label{sub:Evol_stat}

We explore the temperature range from T=25 to T=200 MK with  a step of 5 MK \footnote{In PCI07 we explored only the domain between 25 and 80~MK, since at that time no data were available for chemical elements like K that imply H-burning at much higher temperature.}. We keep always the density at $\rho$=10 \gcc \, which is representative of  typical average values in the cores of massive stars and in the shells of AGB stars, the two main candidate sites for the composition of the 2P stars in GCs.
Since timescales vary a lot as function of temperature, we display our results as a function of the consumed H mass fraction (the initial  one being X$_{\rm H}$=0.75).
 
\begin{figure}
\begin{center}
\includegraphics[width=0.49\textwidth]{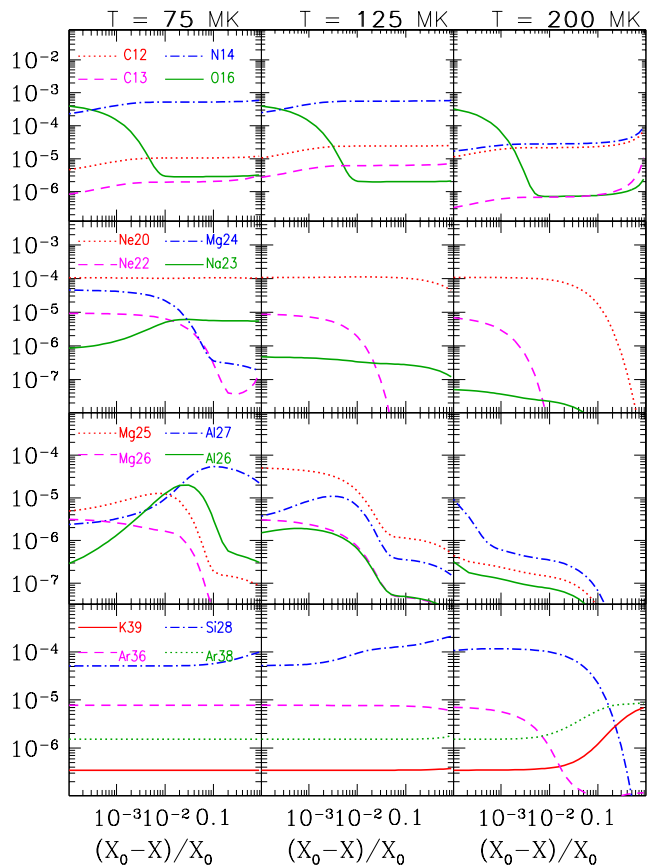}
\caption[]{Evolution of the abundances (by mass fraction) of various elements (as indicated in the leftmost panels) as a function of the consumed hydrogen fraction (used as a proxy for time), for three different values of  constant temperature (as indicated in the top of the figure) and for a constant density of $\rho$=10 \gcc.  
}
\label{Fig:XvsH_stat}
\end{center}
\end{figure}
 
Results of our calculations for three different values of the temperature, appear in Fig.~\ref{Fig:XvsH_stat}. The abundances of selected C, N, O, Ne, Na, Mg, Al and Si, Ar, and K isotopes
are plotted as a function of consumed mass fraction of hydrogen $\Delta {\rm X}/ {\rm X}_{\rm 0}$= (X$_0$-X)/X$_0$. 

In the top row it is seen that the CN isotopes reach rapidly their equilibrium values,  on a timescale shorter than the leftmost value on the time axis of the figure
\footnote{For that reason, the initial values of C and N isotopes do not appear on the figure. By mass fraction, they are 1.3 10$^{-4}$, 6. 10$^{-7}$, 6.2 10$^{-5}$ and 3.1 10$^{-7}$ for $^{12}$C, 
$^{13}$C, $^{14}$N and $^{15}$N, respectively.}, while O reaches equilibrium when $\sim$1 \% of H is burned.  All these isotopes keep their equilibrium values up to H exhaustion for the whole range of temperatures explored here. The carbon isotopic ratio in particular decreases from the assumed initial value of 210 to low values between 3.7 and 30 respectively for burning temperatures of 75 and 200~MK.

$^{23}$Na (2nd row) is produced through the destruction 
of $^{22}$Ne already at 25 MK (not shown here).
As temperature increases, an increasing fraction of $^{20}$Ne  is converted 
to $^{22}$Ne and $^{23}$Na. 
The equilibrium value of $^{23}$Na is highest around T=50 MK and decreases slowly 
at higher T; at 100 MK the equilibrium value is smaller than the initial one, i.e. $^{23}$Na 
is destroyed.   

\begin{figure*}
\begin{center}
\includegraphics[angle=-90,width=0.99\textwidth]{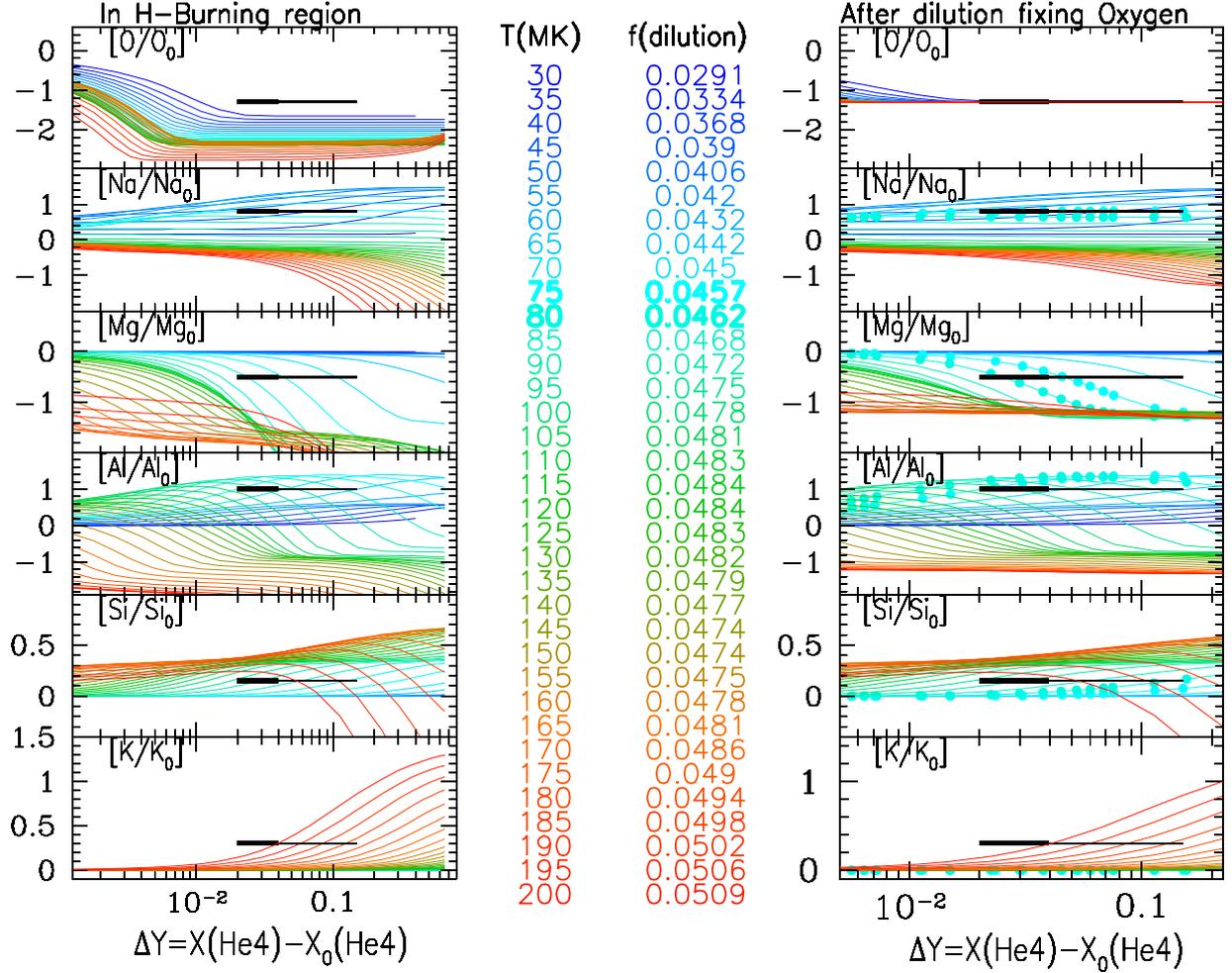}
\caption[]{{\it Left}: Evolution of the abundances (by mass fraction) of various elements as a function of increase in He mass fraction $\Delta$Y (used as a proxy for time), for  different values of  constant temperature (as indicated by the color coded values in the middle column) and for a constant density of $\rho$=10 \gcc; abundances are now expressed as  [X/X$_0$]=log(X)-log(X$_0)$, where X$_0$ is the initial one. The horizontal bars in each panel indicate the differences between the extreme abundances ($\Delta$[X/Fe]=[X/Fe]$_E$-[X/Fe]$_0$) observed in NGC~2808 (see Table 1). Their thick portion extends between  \DY=0.02 and 0.04, variation observed in most globular clusters 
while the  thin portion extends up to $\Delta Y$=0.15, derived for NGC~2808 (see text for references). {\it Right}: Same as on the left, but this time the equilibrium abundances of Oxygen (which displays the largest depletion with respect to its initial value) are diluted by a factor $f$ with pristine material, as to fix the observed extreme value in NGC~2808 for each temperature. That factor $f$, appearing in the middle column for each temperature,  is then used to dilute all other abundances. Points are used for temperatures of 75 and 80 MK, where the curves for Na, Mg and Al match the observations (horizontal bars)
}
\label{Fig:XvsH_new}
\end{center}
\end{figure*}

$^{24}$Mg (3d row) is affected only above 70 MK, and only 1\% of its initial abundance is left at T=80 MK. Early on (small H fractions consumed), its destruction leads to an increase of the abundances of \mgb \ and $^{26}$Al. When more H is consumed, almost all initial Mg (all three isotopes) has turned into $^{27}$Al. 
At temperatures higher than 100 MK, $^{27}$Al is destroyed quite rapidly, the amount of consumed H decreasing with increasing temperature.  $^{27}$Al undergoes an important overproduction in a limited range of temperatures, around 75 - 90 MK, as already discussed in PCI07.

The bottom panel of Fig. \ref{Fig:XvsH_stat} displays the evolution of  Si, Ar, and K. Above 100 MK, Si is overproduced from the increased leakage out of the Mg-Al cycle that starts at $\sim$ 65~MK \citep[e.g.][]{Arnould99}. Above 180 MK, Si starts being destroyed in its turn. The abundances of the heavier isotopes (Ar and K) are not affected by nuclear reactions until temperatures get above 150 MK, when $^{36}$Ar is depleted to the profit of $^{38}$Ar. Production of   $^{39}$K occurs only above 170 MK, when almost all $^{36}$Ar has gone. $^{39}$K reaches a maximum overabundance at 200 MK, when it shares with $^{38}$Ar almost all the initial abundance of $^{36}$Ar. It displays then an overabundance by a factor of 30 with respect to its initial value.  This is practically the maximum abundance that K may reach in our set-up (i.e. with the assumed initial abundances), because at even higher temperatures it is destroyed by (p,$\gamma$) reactions.

Notice that when K gets to such high abundances, all major lighter species - participating in the Ne-Na and Mg-Al cycles - as well as Si, have essentially disappeared. Moreover, more than 30\% of the initial H has been consumed in that case; this has some important consequences, as will be discussed in the next section. Finally, 
it should be noticed that Ar and K never reach equilibrium values, in contrast to the cases of the isotopes of CNO or Ne-Na cycles.

\section{Comparison to NGC~2808}
\label{sec:CompNGC2808}

In this section, we use the nucleosynthesis calculations of Sec.~\ref{sec:Nucleo} to infer the temperatures in the H-burning zones of the "polluters" that affected the composition of 
the intracluster gas out of which NGC~2808 2P stars formed. In Sec.~\ref{subsec:Comp_extreme}, we use the extreme abundances of Table~1 to infer the range of appropriate temperatures, and we summarize the key nucleosynthesis points in Sec.~\ref{subsec:inferences}. We explore the whole abundance pattern of NGC~2808 in Sec.~\ref{subsec:Comp_all}. 

\subsection{From extreme  to predicted abundances,  after dilution}
\label{subsec:Comp_extreme}

In Fig.~\ref{Fig:XvsH_new} we show the elemental overabundances of O, Na, Mg, Al, Si, and K (from top to bottom) as function of $\Delta \rm Y$, the increase in the He mass fraction, 
for all the temperature values of our calculations.  Results are plotted on the left panel as over(under)abundances [X/X$_{\rm 0}$]
with respect to the initial values of those elements, i.e. those corresponding to NGC~2808 1P stars.
In the logarithmic scale adopted in the figure, the value of 0 for [X/X$_{\rm 0}$] represents the initial value of the element.

The results can be understood in terms of the isotopic abundance evolution already analyzed after Fig.~\ref{Fig:XvsH_stat}, taking into account that e.g. the abundance of Mg in Fig.~\ref{Fig:XvsH_new} (left) is the sum of $^{24}$Mg+$^{25}$Mg+$^{26}$Mg+$^{26}$Al in Fig.~\ref{Fig:XvsH_stat}.
We also indicate with a horizontal bar in Fig.~\ref{Fig:XvsH_new} the  maximum variation (i.e. the difference, positive or negative), from [X/X$_{\rm 0}]$ in the abundance of each of the plotted elements, as observed in NGC~2808 and given in Table~\ref{tab:variationsabundancesNGC2808}. The bar extends from the lowest values \DY=0.02 
to 0.04 derived by multiwavelength photometry in other GCs 
and up to \DY=0.15, which is the mean enrichment value evaluated for NGC~2808 (see references in Sec. 2). 
In the case of oxygen, the extreme observed variation is of -1.3 dex, i.e. O is depleted by a factor of 20 in the most extreme 2P stars of NGC~2808.

For the whole temperature range covered by the computations, the O equilibrium values on the left top panel of Fig. 
\ref{Fig:XvsH_new} are reached at the very beginning of H-burning (i.e., for $\Delta$Y much lower than the maximum value allowed by the observations not only for NGC~2808 but also for the other GCs; moreover, the higher the temperature, the lower the value of $\Delta$Y at which O reaches equilibrium). Additionally, they lay systematically below the observed extreme abundance of 
that element in NGC~2808. This suggests that this extreme abundance (and, by extension, the less extreme ones) were obtained by diluting one part of processed material from the H-burning region with $f$ parts of pristine one. As shown in PCI07 for the case of NGC~6752, the whole abundance pattern of GCs can be interpreted in terms of abundances produced by hydrostatic H-burning in some temperature range, diluted with pristine material (having the same composition as the initial one adopted in the calculations) by various factors. PCI07
adopted a simple formalism for the abundances (mass fractions) of the mixture (=2P stars), namely
\begin{equation}
X_{mixed} \ = \ \frac{X_{processed}+f X_{original}}{1 \ + \ f}
\label{equ:dilution_factor}
\end{equation}
where $f$ is the dilution factor, i.e. one part of processed material is mixed with $f$ parts of unprocessed one. The required {\it minimum} dilution factors $f_{min}$ (i.e. those needed to obtain the observed extreme value of O in NGC~2808) for each temperature are then obtained as
\begin{equation}
f_{min} \ = \ \frac{X_{O,extreme}-X_{O,processed}}{X_{O,original}-X_{O,extreme}}
\end{equation}
where we take the values for $X_{O,extreme}$ and for X$_{O,original}$ from Table~\ref{tab:variationsabundancesNGC2808}. The corresponding values of $f_{min}$ for each temperature appear in the second column of the central part of Fig.~\ref{Fig:XvsH_new}. They are quite small, in the range of 0.03 to 0.05, implying that the observed extreme values of O in NGC~2808 are obtained from material processed at nuclear burning temperatures, which has been little diluted with pristine material. The implications of such a small dilution factor are discussed in Sec.  \ref{subsec:dilution}. 

These dilution factors $f_{min}$ are then applied to all the abundances of the left panels and the results appear on the right panels of Fig.~\ref{Fig:XvsH_new}. All equilibrium values of O correspond now to the extreme values observed in NGC~2808 (by construction). In the case of the other elements, extreme observed values for Na, Mg, and Al are obtained for temperatures in the range 75 to 80 MK (indicated by filled circles on right panels in Fig. 2), as already shown in PCI07. This happens for maximum variations in He mass fraction in agreement with the observational constraints for $\Delta$Y in NGC~2808 (0.15 to 0.2), for the whole temperature range considered here.  
In addition, the higher the temperature, the lowest the variation in He mass fraction at which the predictions match the Mg data. Therefore, in the case of other GCs where $\Delta$Y is determined to be significantly lower than in NGC~2808 (i.e., of the order of 0.02 to 0.04), a temperature of $\sim$ 80~MK is required to fit the observed extreme values of Mg at very low He-enrichment. 

In the 75 to 80 MK temperature range, Si and K variations are negligible.  
In order to obtain Si and K overabundances comparable to the extreme observed ones in NGC~2808, temperatures of $>$100 MK and 190 MK are required, respectively. However, in such high temperatures, Na, Mg and Al are largely depleted. 

\subsection{Implications for dilution }
\label{subsec:dilution}

In the previous section we have shown that O gets rapidly to its equilibrium value, when $\sim$1 \% of H is burned and it is the most affected element, its abundance being reduced by factors of 50-500 in the studied temperature range.  Such low O values are never observed in GCs (although for the most O-poor stars data analysis provides only upper limits for this element), implying that some mixing of burned with pristine material has taken place before the formation of 2P stars. The corresponding dilution factor is determined by the observed extreme O abundance and it is generally quite small, of the order of a few \%.

PCI07 found that  in the case of NGC~6752, where  the observed lowest O abundance  is only $\sim$5 times smaller than the pristine one, the corresponding minimal dilution factor is $f_{min}\sim$30 \%, i.e.  considerably higher than in the case of NGC~2808. 

The minimal dilution factor $f_{min}$ found for NGC~2808 is extremely constraining for all scenarios  of 2P star formation. It implies that the extreme 2P stars of NGC~2808 are made  from H-processed material  essentially unmixed with the gas from which the 1P stars were made. It also has a clear implication concerning the presence of a fragile element, like Li. In the case of NGC6752, \citet{PCI07} showed that the Li abundance of the most Na-rich stars of that cluster (a factor of $\sim$3 below the so-called "Spite-plateau", e.g.,  \citealt{SpiteSpite1982Nature}, \citealt{CharbonnelPrimas05}, \citealt{Melendezetal10}) can be explained by assuming mixture of H-processed Li-free material with Li-normal pristine gas with the derived dilution factor of 30 \% \footnote{The dilution factors we derived in PCI07 and in the present study agree with those derived for the mixture between the
stellar winds and the original matter in the case of the FRMS scenario based on Li observations (e.g. \citealt{Decressin07b}). Although this approach differs from that of \citet{SalarisCassisi14} who study dilution of Li-free polluters ejecta directly within pre-main sequence stars in the early-disc accretion scenario of \citet{Bastianetal2013a}, these authors also conclude for the need to mix polluters yields with gas containing pristine Li.}. 
Now, our smaller dilution factor of $\sim$5 \% for NGC~2808  implies that the initial Li content of the most extreme 2P stars of that cluster has to be even lower, about a factor of 10 below the Spite plateau. We are not aware of Li observations in NGC~2808 as function of e.g. O/Fe or Na/Fe, but we urge such observations, as we find them crucial for the validity of all scenarios proposing 2P star formation from mixing of processed and pristine material in GCs. Such observations should however focus on main sequence stars only, as Li is quickly depleted in subgiant and red giant stars under the effects of the first dredge-up and other mixing processes. Even in the case of main sequence stars actually, the interpretation of the Li observations in GCs requires great caution, as its surface abundance may vary under the action of atomic diffusion combined to rotation-induced turbulent processes (for a review and references, see \citealt{Charbonnel16_ees}).

Finally, the carbon isotopic ratio obtained after dilution with a minimum dilution factor of 0.0455 (as found for H-burning temperature of 75-80 ~MK, see Fig. 2) is $^{12}$C/$^{13}$C=4.4. This is in agreement with the  ratio derived for a handful of subgiant stars in a couple of GCs (NGC~6752 and 47~Tuc; \citealt{Carrettaetal05}) but not yet in NGC~2808. As discussed by \citet{Charbonnel14}, the determination of this quantity in turnoff GC stars could provide a new signature of membership to the first and second stellar populations.

\subsection{Inferences on  H-burning temperature}
\label{subsec:inferences}

\noindent
Applying the aforementioned  mixing/dilution factor $f_{min}$ to all other elements (less affected than O), we find that:

\noindent
-Observed extreme values for Na, Mg, and Al in NGC~2808 can be obtained for material that has been processed in the temperature range 75-80~MK, in agreement with PCI07 finding for NGC~6752.

\noindent
- The corresponding predicted He enrichment agrees with the observational constraints on $\Delta$Y for NGC~2808. Lower values of $\Delta$Y as observed in other GCs require temperature of the order of 80~MK to fill the Mg constraint.

\noindent
- For extreme overabundances of Si and K, temperatures higher than 100 and 190 MK, respectively are required, but then Na, Mg and Al are destroyed. This implies that {\it{high K abundances are not expected to be seen in Na- and Al-enriched stars in the case of a single nucleosynthesis source for GC abundance anomalies.}}

\begin{figure}
\begin{center}
\includegraphics[width=0.49\textwidth]{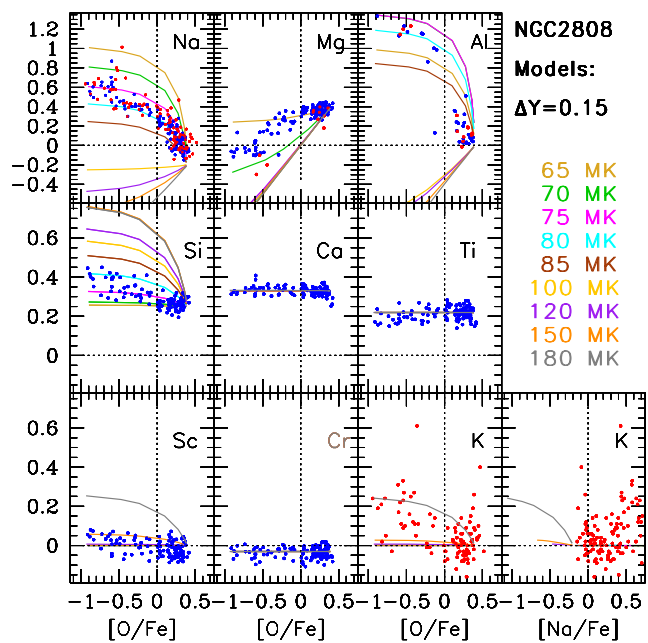}
\caption[]{The theoretical abundances of various elements of Fig.~\ref{Fig:XvsH_new} obtained at \DY=0.15  are diluted by various amounts of pristine material and plotted as function of the corresponding values of [O/Fe] for selected values of temperature (color-coded lines). Data for NGC~2808 are from \citet[][blue]{Carretta14,Carretta15} and \citet[][red]{Mucciarellietal15}. Typical error bars are of the order of 0.1~dex as quoted in the original papers
}
\label{Fig:CompDataAllElements}
\end{center}
\end{figure}


\subsection{The abundance pattern of NGC~2808}
\label{subsec:Comp_all}

 In the previous section, we determined the factor $f_{min}$ required to explain the extreme abundances of NGC~2808 for O and found it to be of a few \%. 

We shall use this simple scheme here,    assuming that $X_{processed}$  for each temperature is the value obtained in the simulations when \DY=0.15 (see Fig. \ref{Fig:XvsH_new}). This choice automatically ensures that i) O will be at its equilibrium value, so we can start dilution with the $f_{min}$ values of the previous paragraph and ii) the corresponding He abundance will be the extreme one observed in NGC~2808.  

In Fig.~\ref{Fig:CompDataAllElements} we display the results of Na, Mg, Al, Si, Ca, Ti, Sc, Cr, and K as function of O, as well as K vs Na, for various values of temperature and for increasing larger values of the dilution factor $f$, starting with $f_{min}$. We compare our predictions with NGC~2808 observations \citep{Carretta14,Carretta15,Mucciarellietal15}. Since our step in temperature is of 5~MK, we can only perform a coarse-grained analysis of the situation; a smaller step would be required for a finer analysis.

Na and Al patterns are correctly reproduced for T=75 MK, but Mg fits better at T=70 MK and disappears above that temperature, when \DY=0.15. The Si pattern fits best at T=80 MK. Taking into account observational error bars, of the order of 0.1 dex in most cases   we find 
that the temperature range of 70-80 MK is the one best reproducing the aforementioned patterns. This is similar to our conclusion in PCI07, which concerned NGC~6752, while here we have also included Si and the \DY \ constraint.

We find that the K vs O pattern can be obtained for T$\ge$180 MK in those conditions. However, at that temperature Na and most other elements have already been depleted to very low values and {\it{the K-Na correlation observed in NGC~2808 is not predicted at any temperature}}. This inconsistency is obtained whatever the He abundance variations, i.e., independently of the He constraint. Similarly, the computations predict a K-Al anticorrelation, at odds with the data.

\begin{figure*}
\begin{center}
\includegraphics[angle=-90,width=\textwidth]
{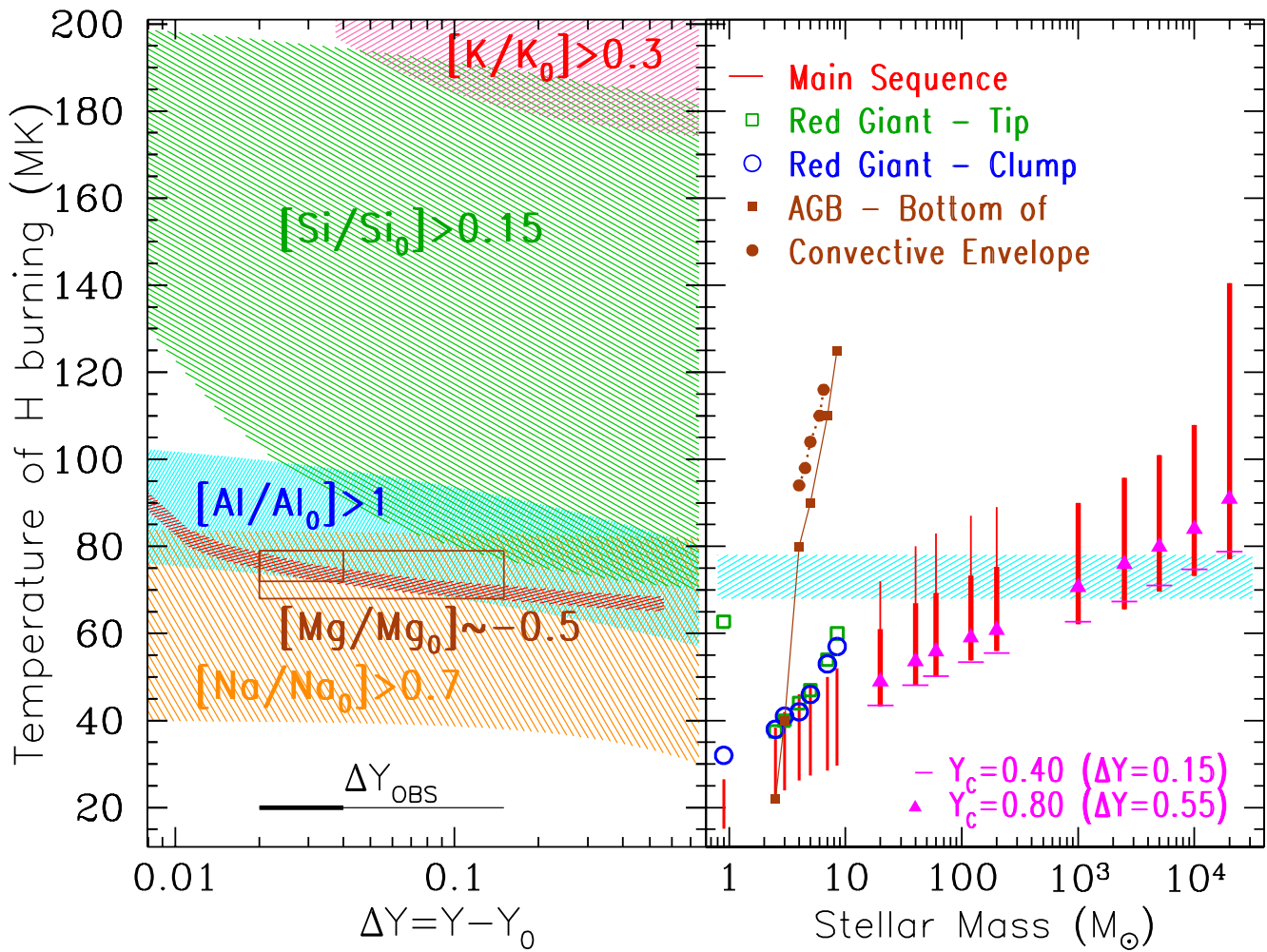}
\caption[]{{\it Left:} Shaded aereas display the temperature range as a function of produced He amount  ($\Delta \rm Y$) where each one of the corresponding elements  is either overproduced (Na, Al, Si and K) or depleted (Mg) with respect to its initial value by the factors indicated;  these factors correspond to the extreme abundances observed in NGC~2808, as given in Table 1.  The horizontal bar in the bottom corresponds to the observationally inferred variation in the He content of most globular clusters (thick portion) and of NGC~2808 (thick portion), see Sec. 2. The boxes around the Mg values correspond to those inferred variations and show explicitly why {\it Mg is the most sensitive "thermometer" among those elements }(see text).
{\it Right:} H-burning temperatures at various phases of a star's life, as a function of 
initial stellar mass. The correspondence between symbols and evolutionary phases 
is displayed in the top of the figure (adapted from Prantzos et al. 2007). 
Vertical red segments (for central H-burning) start at the ZAMS and end at central H mass fraction X=0.01 (the thick portion) and X=0 (the thin portion, for stars between 20 and 200 \ms). For the other phases we show only the maximum temperature reached in the H-burning zones of interest. In the bottom part, more details are given for massive star evolution.
On the rightmost part of the figure, the temperatures are for the supermassive star models of Denissenkov (private communication). The shaded rectangular aerea indicates the "triple point", i.e. the temperature range (from the left part) allowing for the co-production of Na, Mg and Al at the levels inferred from observations of NGC~2808. 
}
\label{Fig:TvsMandY}
\end{center}
\end{figure*}


\section{Implications for the polluter candidates}
\label{sec:Polluters}

\subsection {T vs $\Delta$Y constraints for hydrostatic H-burning  }
\label{sec:TemperatureRange}

In Sec.~\ref{subsec:Comp_extreme}  we  determined the temperature conditions allowing for the production/destruction of the key elements O, Na, Al, Mg, Si, and K to their extreme abundance levels, as observed in NGC~2808 and considering the He constraint. In Sec.~\ref{subsec:Comp_all} we showed that, within a simple scheme of dilution of those abundances with pristine material, one may reproduce a large amount of the observed patterns in NGC~2808. However, even if some correlations/anti-correlations are satisfactorily reproduced, this does not always occur for the same range of temperatures. Furthermore, some observed correlations (K-Na and K-Al) are never reproduced.  

The situation is summarized on the left part of Fig.~\ref{Fig:TvsMandY}. 
It displays the temperature ranges where each of the key elements reaches (after dilution) its extreme abundance, as observed in NGC~2808 and indicated by the corresponding numbers. The temperature ranges are displayed as a function of the increase \DY \ in He mass fraction. We indicate the observationally inferred variation in He content for NGC~2808 as well for other GCs. It can be seen that Na and Al "coexist" at their observed highest level in a narrow range around T$\sim$75-80 MK at low $\Delta$Y (compatible both with the relatively high He values derived for NGC~2808 and 
with the much lower He values obtained in several GCs), which extends down to 60-80 MK at H-exhaustion ($\Delta$Y $\sim$0.75). Na drops below its extreme value above T$\sim$80 MK, and Al above 80-90 MK, depending on He amount. At low \DY \ values, Si is produced after Al decreases from its maximum abundance.
For larger $\Delta$Y, Si may coexist at its observed level with large Al amounts around 90 MK and even with Na around 80 MK, but only for $\Delta$Y values larger than derived observationally in NGC~2808. K is produced for relatively high $\Delta$Y (still compatible with $\Delta$Y in NGC~2808, but not with $\Delta$Y $<$0.04). However and as already emphasized before, K production occurs only at extremely high temperatures where Na, Al, and Mg are destroyed; it is marginally compatible with Si enhancement for very high $\Delta$Y beyond the observational constraint. 

An interesting feature is the very narrow temperature range where Mg displays its observed depletion of [Mg/Mg$_0$]$\sim$-0.5~dex in NGC~2808: a narrow band of width $\Delta \rm T \sim$5 MK extending from T$\sim$80 MK at $\Delta$Y=0.02 to   T$\sim$70 MK at $\Delta$Y=0.15. In PIC2007 we had shown that not only Mg, but also its isotopic abundances are co-produced at the levels observed in NGC~6752 for T$\sim$75 MK, but for smaller \DY \ values.  {\it The derived narrow temperature range suggests that the abundance of Mg constitutes the most sensitive thermometer to probe the conditions of the astrophysical site of the polluter(s); the increase of the He abundance constitutes a "second parameter" in that respect. This is a key finding of our study}. 

\subsection{Hydrostatic H-burning sources}
\label{subsec:HydroHburn}
	
\subsubsection{H-burning temperature in stellar models}
\label{subsubsec:stellarmodels}

We now check whether the aforementioned nucleosynthetic requirements for NGC~2808 correspond to the temperature conditions of hydrostatic H-burning in specific astrophysical sites. We consider single stars in the whole range of masses between 1 and 200 \ms \ (while in PCI07 we considered stars up to 120 \ms ~only), as well as the putative supermassive stars above 10$^3$ \ms that have been proposed as polluters of GCS \citep{DenissenkovHartwick14}. We also consider H-burning shells of red giants and AGB stars. 

The results are summarized on the right panel of  Fig.~\ref{Fig:TvsMandY} where we plot, on the same scale as on the left, the temperatures of the various putative polluters as function of the initial stellar mass. We collect predictions from stellar model computed by various authors at [Fe/H] = -1.5.  The models with masses between 20 and 120 \ms  ~are from \citet{Decressin07b}, and the 200 \ms ~model was computed with the same version of the Geneva stellar evolution code for the present study (C.Georgy, private communication). 
The models with masses lower than 9 \ms ~are a combination from \citet{Decressin09a}, 
\citet{Pumoetal07}, and \citet{Ventura02}. 
Finally, the models for supermassive stars have been computed by P.Denissenkov (private communication) with the stellar evolution code MESA \citep{Paxtonetal11,Paxtonetal13} using the same physical assumptions as in \citet{Denissenkovetal15}.

The correspondence between symbols and evolutionary phases is displayed in the upper left part of the figure (adapted from PCI07). Vertical red segments (for central H-burning) start at the ZAMS and end at central H-exhaustion. We also indicate with small horizontal tick marks the temperatures where \DY $\sim$0.15 (i.e. maximum He enrichment in NGC~2808) and with triangles the temperatures where \DY $\sim$0.55 
(i.e. 20\% of H left). For the other phases we show only the maximum temperature reached in the H-burning zones of interest (i.e., H-burning shell for stars at the RGB tip and at the clump, and base of the convective envelope for stars undergoing thermal pulses on the AGB). 

A comparison of the left and right diagrams of Fig.~\ref{Fig:TvsMandY} allows one to draw the following straightforward conclusions for the different potential polluters\footnote{We do not discuss the case of intermediate-mass binary stars \citep{DeMink09} since the temperature in these 10 - 20 ~M$_{\odot}$ objects can account only for Na-enrichment, but not for Al enhancement or Mg depletion.}. Here we focus on the nucleosynthesis aspects. We refer to recent reviews for an extended discussion about  more general issues related to the associated GC pollution scenarios 
\citep[e.g.][]{Renzinietal15,Bastian_iau316,Charbonnel16_ees,Charbonnel_iaus316}. 

\subsubsection{Fast rotating massive main sequence stars}
\label{subsubsec:massivestars}

Our computations confirm that central H-burning in massive stars (20 to 200 \ms) can produce the observed large Na excess and O depletion (i.e. the defining chemical feature of GC stars) for relatively low \DY \ values, corresponding to the range observed in GC 2P stars (i.e., between 0.01 and 0.2).
In contrast, the observed large Al values and the observed Mg depletion can be obtained for massive main sequence star temperatures only for large increases of the He content (i.e. towards H-exhaustion), substantially larger than inferred from GC observations.
A similar conclusion holds for Si: large amounts, i.e. comparable to observed ones, can be obtained at T$\geq$80 MK but only for high He abundances; however, at such high temperatures Mg is always destroyed. Finally, no K can be produced in these objects. 

Our results agree with the nucleosynthesis predictions of the evolution models of fast rotating massive stars (FRMS, see e.g. Fig.~3 and 6 of \citealt{Decressin07b}). 
In the original FRMS scenario, the hydrogen-burning ashes would be ejected from massive stars rotating near critical rotation velocity {\it{all along the main sequence}} and would mix with pristine gas to form 2P stars in  the  immediate surroundings of the massive parent star.  
Since the CNO-cycle and the NeNa-chain fully operate at relatively low temperature at the beginning of the main sequence, the FRMS models naturally explain the O-Na anticorrelation for moderate He enrichment as derived for NGC~2808 or even lower (e.g. \citealt{Chantereau15}). Additionally, 2P stars are predicted to be born with relatively low carbon isotopic ratios \citep{Charbonnel14}.

However,
2P GC stars  with important Al excess and Mg depletion can be made  only from high temperature material of FRMS close to H-exhaustion,  having therefore a high He mass fraction (up to Y=0.8 after dilution with pristine gas; see e.g. Fig.~1 of \citealt{Chantereau16}). 
Here, we confirm this conclusion. In particular, the difficulty to reproduce the Mg-Al anticorrelation was already pointed out by \cite{Decressin07b} who 
showed that in the typical case of NGC~6752, this pattern could be reproduced by FRMS models only by invoking an increase of the $^{24}$Mg(p,$\gamma$) reaction rate by a factor of 1000 around 50MK with respect to the nominal rate of \citet{Iliadisetal01}.
Such an increase is extremely unlikely, given the quality of current input nuclear physics information for this rate.
Even in that case, Mg-depletion would come with strong He enrichment. 

Whether or not this could be detected in the GC CMDs remains unclear. In the original FRMS scenario indeed, only $\sim 20\%$ of 2P in NGC~2808 are expected to be born with an initial \DY ~higher than 0.15 \citep{Chantereau16}. Because of
the effects of He on the stellar evolution paths and lifetimes \citep[][and references therein]{Chantereau15}, their total number reduces to $\sim 10\%$ at the main sequence turnoff and on the RGB for GCs with ages between 9 and 13~Gyr.  Additionnally, for the same age range, this fraction
drops dramatically on the horizontal branch ($\sim 0.5\%$ at 13~Gyr), meaning that this evolution phase cannot be used to distinguish between the two most commonly invoked scenarios for
GC enrichment. These very helium-rich 2P stars might even disappear on the AGB, depending on the age and metallicity of their host GC, as well as on their mass loss rate along the RGB \citep{Charbonnel13,Cassisietal2014,CharbonnelChantereau16,Chantereau16}.

\subsubsection{Super-massive main sequence stars}
\label{subsubsec;sms}

The situation appears to be more favorable in the case of supermassive main sequence stars (SMS) with masses above $2 \times 10^3$ \ms.
According to the stellar models we plot in Fig.~\ref{Fig:TvsMandY}, their central temperatures are already high enough at the beginning of the main sequence to allow for a simultaneous co-existence of He, Na, Al and Mg to levels compatible with observational constraints (except for SMS with mass M$>$10$^4$ \ms). For these elements, our computations are in good agreement with the model predictions of \citet[][see e.g. their Fig.~1 for maximum $\Delta$ Y of 0.15]{Denissenkovetal15}. Our computations show additionally that
Si can also be produced in that site, but at temperatures where Mg is heavily depleted. When both constraints are considered, i.e., T and \DY, we find that neither Mg nor Na survive Si production. Finally, we show that the current SMS models do not reach K-production temperature.

As can be seen in Fig.~\ref{Fig:TvsMandY}, the models predict a strong increase of the central temperature as SMS evolve along the main sequence, in particular once \DY ~reaches 0.55 (central He mass fraction of 0.8), which leads to Na and Al depletion, at odds with the observations. Therefore, if SMS played a role for GC pollution, they must have released  H-burning material at the very beginning of the main sequence, as proposed by \citet{DenissenkovHartwick14} based on the He argument. Whether this can happen requires the verification of speculations on how the required mass-loss might be driven by the super-Eddington radiation continuum-driven stellar wind or by the diffusive mode of the Jeans instability.

\subsubsection{Asymptotic giant branch stars}
\label{subsubsec;agb}

For the case of AGB stars, conclusions are not straightforward, because the simple 1-zone calculations at constant T performed here cannot capture the complexity of the situation: In addition to hot-bottom burning in the AGB envelope that is invoked for the required hot H-burning, 3d dredge-up events can occur and mix material of He-burning to H-burning zones, altering the composition of the latter and the products of nucleosynthesis \citep[for details see e.g.][]{Charbonnel16_ees}. The range of AGB temperatures reported in Fig. \ref{Fig:XvsH_new} thus simply indicates that the abundances of Na, Mg, Al, O, and Si may be modified in that site, while K can not, but it gives no clues as to the resulting correlations or anti-correlations between those elements, which are strongly determined by the mixing with He-burning material. 
For example, all the AGB and super-AGB models computed so far predict chemical yields at odds with the observed O-Na anticorrelation \citep[e.g.][]{Forestini97,DenissenkovHerwig03,KarakasLattanzio07,Siess10,Ventura13,Dohertyetal14II}, exactly because of the competition between 3d dredge up and HBB \citep[for details see e.g.][]{Charbonnel16_ees}.

Moreover, models of rotating AGB stars predict the ejection of helium-burning products by these stars \citep{Decressin09a}. During central He-burning indeed, rotational mixing brings fresh CO-rich material from the core towards the hydrogen-burning shell, causing important production of primary $^{14}$N that is later mixed in the stellar envelope during the 2DUP. This strongly increases  the total C+N+O content at the stellar surface and in the stellar wind (this adds to the C+N+O increase already predicted by standard, non-rotating models, see e.g. \citealt{Karakasetal06He}).
While this prediction is sustained by a wide variety of galactic observations \citep[e.g.][]{Maeder06,Chiappinietal08}, 
in the case of GCs it would imply that 2P stars present higher total C+N+O values than their 1P counterparts. This at odds with the constancy of C+N+O observed among GC low-mass stars which otherwise present strong variations in the individual abundances of C, N, O, and Na (references in \S~\ref{sec:Intro}).

The temperature range for hot-bottom burning by the AGB models shown in Fig.~\ref{Fig:TvsMandY} does not reach the values that are necessary to produce K. This is totally consistent with the results of \citet{Ventura12} when they use the current nuclear reaction rates available for the reactions involved in the production of this element. These authors thus call for an increase of the $^{38}$Ar(p,$\gamma ^{39}$)K cross section by a factor of 100 to produce K in AGB models. However, an increase of this magnitude seems unlikely, considering the nuclear physics input.
 As an alternative,  \citet{Ventura12} suggest that HBB temperature might be higher in AGB stars than the current models predict. If this were to be the case, the computations presented in the present paper show that the corresponding yields would then be devoid of Na, Al, and Mg, as discussed previously. 

Finally, our computations do not provide any information on the amount of He that is produced by AGB stars and released in their winds together with the H- and He-burning products. In the case of these potential polluters indeed,  He enrichment comes from the 2d dredge-up that occurs on the early-AGB phase (He enrichment due to HBB is negligible). As a consequence, the He yields on one hand and those of CNONaMgAl nuclei are not directly correlated, since these elements are processed at different phases of the evolution. In all the AGB models available (references above), the He content of the ejecta increases with polluters mass, and Y reaches a value of $\sim$ 0.38 for the most massive AGBs. All 2P stars are thus expected to share similar and relatively high initial He abundances, independently of their CNONaMgAl content.

\subsection{The peculiar case of K: a hint for novae?}
\label{subsec:Kproduction}

Among the various abundance patterns 
reported so far for GC stars, the overabundance of potassium in the most massive GCs like NGC~2808, if real, is undoubtedly the most disturbing. None of the known or putative stellar sources discussed in the previous section are expected to reach H-burning temperatures as high as $>$180 MK, that are required to  produce K in excess.

H-burning as such high - or even higher - temperatures is expected to occur in novae, but explosively, i.e.  on a hydrodynamical timescale $\tau=\rho^{-1/2}$ (where $\rho$ is the peak density of the exploding region) which is always shorter than the nuclear timescale; as a result, only a small fraction of the main nuclear fuel is consumed. Still, important modifications to the abundances of other nuclear species may occur.

Detailed theoretical investigations have shown that the nova outburst is a multi-parameter phenomenon, depending on e.g. the mass, temperature and composition of the exploding white dwarf, the accretion rate and composition of the material of the companion star etc., see e.g. \cite{Jose2017} for a	 recent review. In those conditions, it is not easy to perform parametrized simulations of nova nucleosynthesis. Still, detailed 1D simulations clearly suggest that peak temperatures much higher than 200 MK, reaching sometimes 450 MK can be obtained \citep[e.g.][]{Yaron2005,Denissenkov2014}.

In particular, \cite{Denissenkov2014} studied models of CO and ONe white dwarfs with the MESA  code, performing post-processing nucleosynthesis with the NUGRID multi-zone code MPPNP. They used a network of 147 species, up to Ca-47  and thus appropriate for K production; 1700 nuclear reactions are considered, with rates taken for JINA Reaclib v1, which include those evaluated by \cite{Iliadis2010}.

An inspection of the results presented in \cite{Denissenkov2014} shows that a modest overabundance of K, by factors of a few, can be obtained in some cases. However, in all those cases, all intermediate mass elements (Na, Al, Si, S) are systematically overproduced to large amounts, much larger than observed; sometimes O is also overproduced, in contrast with its well known depletion in GCs. The reasons is that  such elements are made from abundant "seed" material (CO or ONe white dwarfs) and explosive nucleosynthesis has little time to deplete substantially the original elements. Based on those results, one is tempted to conclude that "K overproduction in novae is expected to be accompanied by much larger overproduction of lighter elements". Such a conclusion would be in stark contrast with the one on hydrostatic H-burning presented in Sec.  4.2. However, a much more thorough investigation of the large parameter space of novae would be required to establish such a conclusion on a firm basis.

The issue of novae as potential sources of K anomalies in the 2G stars of GCs is also discussed in \cite{Iliadisetal16} from the point of view of the total mass available for star formation in that case. They find that for the case of NGC~2419, novae could eject enriched material to form at most  1 \% of the K-enriched stars observed in that cluster (which constitute about 30\% of the total population). But they point out that
current models of novae, concerning white dwarfs accreting from their companion stars, are perhaps not appropriate to account for the situation in the early life of GCs, where accretion would take place directly from the fairly dense ($\sim$10$^6$ cm$^{-3}$) ISM; the total amount of material processed by the white dwarfs of the GC could be much larger in that case. However, the corresponding hydrodynamical problems, concerning the retention of the fast ejecta within the cluster in order to form the 2G stars, have not been considered up to now.


\section{Summary}
\label{section:summary}

In this work, we reassess the nucleosynthesis constraints on the putative polluters of 2P stars in GCs. We consider the specific case of NGC~2808, where abundances for  elements up to K have been reported to display variations among the cluster stars.
We study H-burning nucleosynthesis in hydrostatic conditions, as appropriate for almost all polluter sites proposed so far. We perform parametrized calculations, varying the temperature in the range of 20 to 200 MK and keeping the density constant at 10 \gcc.

In the spirit of our previous study (\citealt{PCI07}, where we focused on NGC~6752) we seek first to determine conditions reproducing the most extreme abundances reported in one specific GC, namely NGC~2808. We find that it is possible to reproduce the extreme abundances of O, Na, Al and Mg for T$\sim$75-80 MK, after a small dilution with pristine material (i.e. having the composition of the 1P stars of the cluster). The small dilution factor, of the order of $\sim$5 \%, is determined by the abundance of oxygen, which is the most affected heavy element in H-burning. This minimal dilution factor is substantially smaller than determined in PCI07 for NGC~6752 ($\sim$30 \%). This finding is extremely constraining for all scenarios of 2P star formation in GCs, which must account for stochasticity in the formation process of MSPs \citep{Bastianetal15}. Here it implies that the extreme 2P stars of NGC~2808 are made from H-processed material essentially unmixed with the gas from which the 1P stars were made. A direct consequence of that is that the initial Li content of the most extreme 2P stars of that cluster has to be about a factor of 10 below the Spite plateau. We urge then Li observations in NGC~2808 main sequence stars, as we find them crucial for the validity of all scenarios proposing 2P star formation from mixing of processed and pristine material in GCs.

The narrow temperature range between 70-80 MK where  the observed extreme values of O, Na, Mg and Al in NGC~2808 can be reproduced (after minimal dilution) is the same as the one determined in PCI07 for NGC~6752. Here we show that the abundance of Mg constitutes the most sensitive thermometer, allowing one  to probe the conditions of the astrophysical site of the polluter(s). The increase of the He abundance constitutes a "second parameter" in that respect, allowing for the same abundance of Mg to be obtained either at T$\sim$80 MK and low \DY \ (a few \%) or at lower T$\sim$70 MK and higher \DY \ (around 10 - 15 \%). Finally, we find that the observed extreme Si abundance in NGC~2808 can also be obtained for the high T values of the aforementioned temperature range, around T$\sim$80 MK, and for high values of \DY $\sim$0.15. 

 Ideally, our scheme should allow us to pinpoint more precisely the allowed region in the T vs \DY \ phase space. However, the uncertainties in the observed abundances (of $\sim$ 0.1~dex) and in the He content of GC stars, preclude at present that possibility. Moreover, the simplicity of the method - 1-zone H-burning, not allowing for replenishment of depleted species through convection - makes it difficult to accurately simulate realistic sites, like  convective core burning in massive stars or hot bottom burning in AGBs. Despite that, the resulting abundances are more sensitive to the variations of temperature, thus our results should be considered as reliable first order  approximations of more complex situations.

 Once the extreme values of the various elements are obtained (after determining T and $f_{min}$), dilution with various factors higher then $f_{min}$ allows one to obtain the full abundance pattern of a given GC, as shown in PCI07. Here we repeat this "trivial" exercise for NGC~2808, showing how most of the observed abundance patterns can be satisfactorily obtained from material processed through H-burning at T$\sim$70-80 MK. K is a clear exception to that, since its observed extreme abundance requires T$\sim$180 MK at least. In such high temperatures, most of the other key elements (Na, Mg,  Al) are heavily depleted and the corresponding patterns do not match the observed ones.  

Finally, we discuss the main putative polluters of GCs, namely AGBs, FRMS, and SMS, considering only the nucleosynthesis aspects and not other features of the corresponding scenarios (e.g. the mass budget problem, the retention of the polluter ejecta or their mixing with pristine material, the dependence on the mass, compactness, and metallicity of the cluster). As already emphasized in the literature, AGBs produce correlations rather than the observed anti-correlations between O and Na, Al. FRMS easily make the O-Na extreme abundances but barely reach the temperature regime for the extreme Mg-Al abundances observed in NGC~2808 as well as in other very massive or metal-poor GC. The latter regime is reached by the hypothetical SMS  resulting in severe depletion of Mg already early on the main sequence, i.e., for low \DY . 
None of the aforementioned sources reaches the high T regime required for K production. The only known H-burning site able to do so is novae, but recent models show that abundance patterns substantially different than the observed ones should be expected in that case. 
So far, K abundance variations have been reported only in two of the most massive galactic GCs. If confirmed, these observations could eventually indicate that specific conditions were ruling the early evolution of such clusters. However, {\it{the fact that K and Na are observed to be correlated can not be explained with our current understanding of H-burning nucleosynthesis.}} Before turning to exotic solutions, we urge observers to investigate possible issues in abundance analysis.

Our conclusion
is that, despite two decades of intense theoretical and observational investigation, the origin of the observed abundance "anomalies" in GCs remains a fundamentally unsettled issue. However, we think that this work convincingly shows that respecting the nucleosynthetic constraints must be the fundamental consideration for all the scenarios that aim to explain the formation of multiple stellar populations in GCs.
Although we focussed here on the specific case of NGC~2808, one of the most massive GCs that show the most extreme abundance anomalies, our computations can be directly used to constrain the polluter stars that may have played a role in shaping differently (or stochastically) the abundance patterns in other GCs.

\begin{acknowledgements}  
We warmly thank P.Denissenkov and C.Georgy for providing stellar models results, and N.Bastian, W.Chantereau, M.Gieles, M.Krause, and C.Lardo for careful reading and useful comments on the manuscript. CC acknowledges support from the Swiss National Science Foundation (SNF) for the Projects 200020-159543 “Multiple stellar populations
in  massive  star  clusters  –  Formation,  evolution,  dynamics,  impact  on  galactic
evolution”  and 200020-169125 "Globular cluster archeology". We thank the International Space Science Institute (ISSI, Bern, CH) for welcoming the activities of the Team 271 "Massive Star Clusters Across the Hubble Time" (2013 - 2016; team leader CC).
\end{acknowledgements}

\bibliographystyle{aa}
\bibliography{Reference}

\end{document}